# The THESEUS space mission: updated design, profile and expected performances

L. Amati[a], P. O'Brien [b], D. Götz [c], E. Bozzo [d], A. Santangelo[e,] on behalf of the THESEUS Consortium[1]

[a]INAF - OAS Bologna, via P. Gobetti 101, I40129 Bologna, Italy;
[b] Department of Physics and Astronomy, University of Leicester, Leicester LE1 7RH, UK
[c] IRFU/Département d'Astrophysique, CEA, Université Paris-Saclay, F-91191 Gif-sur-Yvette, France
[d] Department of Astronomy, University of Geneva, ch. d'Écogia 16, CH-1290 Versoix, Switzerland
[e] Institut für Astronomie und Astrophysik, Abteilung Hochenergieastrophysik, Kepler Center for Astro and Particle Physics, Eberhard Karls Universität, Sand 1, D 72076 Tübingen, Germany

**ABSTRACT**

THESEUS is a space mission concept, currently under Phase A study by ESA as candidate M5 mission, aiming at exploiting Gamma-Ray Bursts for investigating the early Universe and at providing a substantial advancement of multi-messenger and time-domain astrophysics. In addition to fully exploiting high-redshift GRBs for cosmology (pop-III stars, cosmic re-ionization, SFR and metallicity evolution up to the "cosmic dawn"), THESEUS will allow the identification and study of the electromagnetic counterparts to sources of gravitational waves which will be routinely detected in the late '20s / early '30s by next generation facilities like aLIGO/aVirgo, LISA, KAGRA, and Einstein Telescope (ET), as well as of most classes of X/gamma-ray transient sources, thus providing an ideal sinergy with the large e.m. facilities of the near future like, e.g.,  LSST, ELT, TMT, SKA, CTA, ATHENA. These breakthrough scientific objectives will be achieved by an unprecedented combination of X/gamma-ray monitors, providing the capabilities of detecting and accurately localize and kind of GRBs and may classes of transient in an energy band as large as 0.1 keV - 10 MeV, with an on-board NIR telescope providing  detection, localization (arcsec) and redshift measurement of the NIR counterpart. A Guest Observer programme, further improving the scientific return and community involvement is also envisaged. We summarize the main scientific requirements of the mission and provide an overview of the updated concept, design (instruments and spacecraft) and mission profile.

## 1. INTRODUCTION

The Transient High-Energy Sky and Early Universe Surveyor (THESEUS) is a space mission concept developed by a large European collaboration and submitted in 2016 to the European Space Agency (ESA) in response to the Call for next M5 mission within the Cosmic Vision Programme[1,2]. In 2018, THESEUS, together with other two mission concepts (SPICA and EnVision) was selected by ESA for a 3-years Phase A assessment study, that will end in the first half of 2021 with the Mission Selection Review (MSR) and the down-selection of one candidate in June 2021. The current ESA/M5 schedule foresees q final decision on mission adoption in 2024 and a launch from Kourou in 2032.

---

[1] http://www.isdc.unige.ch/theseus (see Figures captions and Acknowledgments section for specific contributions to this article)

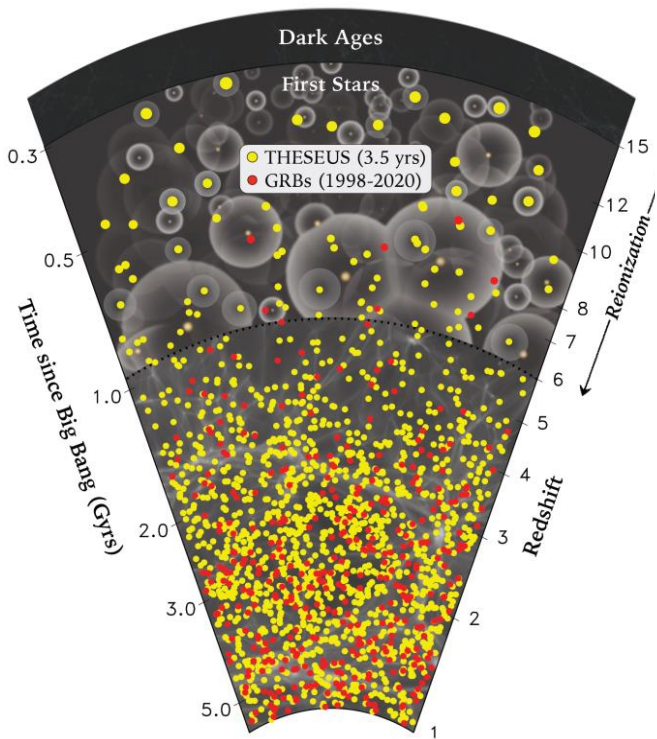

Figure 1 THESEUS capability of detecting and autonomously identifying high-redshift GRBs, as a function of cosmic age, in 4 years of operations (red dots) compared to what has been achieved in the last ~20 years. *(P. Rosati and THESEUS Consortium)*

THESEUS is designed to fully exploit the unique and breakthrough potentialities of Gamma-Ray Bursts (GRB) for investigating the Early Universe and advancing Multi-Messenger Astrophysics, while simultaneously vastly increasing the discovery space of high energy transient phenomena over the entirety of cosmic history[1,2,3]. The primary scientific goals of the mission will address the Early Universe ESA Cosmic Vision theme "How did the Universe originate and what is made of?" and will significantly impact on "The gravitational wave Universe" and "The hot and energetic Universe" themes. These goals will be achieved by a payload and mission profile providing an unprecedented combination of: 1) wide and deep sky monitoring in a very broad energy band (0.3 keV - 10 MeV); 2) focusing capabilities in the soft X-ray band providing large grasp and high angular resolution; 3) on board near-IR capabilities for immediate transient identification, arcsecond localization and redshift determination; 4) a high-degree of spacecraft autonomy and agility, together with capability of promptly transmitting to ground transient trigger information.

By satisfying the requirements coming from the above main science drivers, the THESEUS mission will automatically provide fundamental synergies with the large multi-wavelength and multi-messenger facilities of the future allowing them to fully exploit their scientific capabilities. As outstanding examples, we remark that: a) THESEUS would operate at the same epoch of ATHENA, and would be the ideal space mission to provide the triggers required to fulfil some of the scientific objective of this mission involving GRBs (progenitor environment, find pop-III stars, allow high S/N absorption spectroscopy of Warm-Hot Intergalactic Medium, WHIM) and high-energy transients of all kinds; b) THESEUS will provide high-energy transient survey capabilities complementary to those of the Large Synoptic Survey Telescope (LSST) in the optical; the joint availability of the two facilities in the next decade would provide a substantial advancement of time-domain astronomy. THESEUS will also enable excellent guest observatory science opportunities, including, e.g., performing Near Infra-Red (NIR) observations, long-term monitoring, and will provide capability for rapid response to external triggers, thus allowing strong community involvement.

The data policy for THESEUS will maximize the participation of the international community and minimize cost and complexity of the management by the scientific ground segment. As in the case of the very successful Swift mission, the data will be reserved to the Instrument Teams only during the launch and Early Orbit Phase; data rights will be extended to scientists of the whole consortium during the Performance Verification Phase; and after that, data will be public as soon as they are processed. ESA and the Consortium will do their best efforts for supporting the scientific community in the full exploitation of THESEUS data.

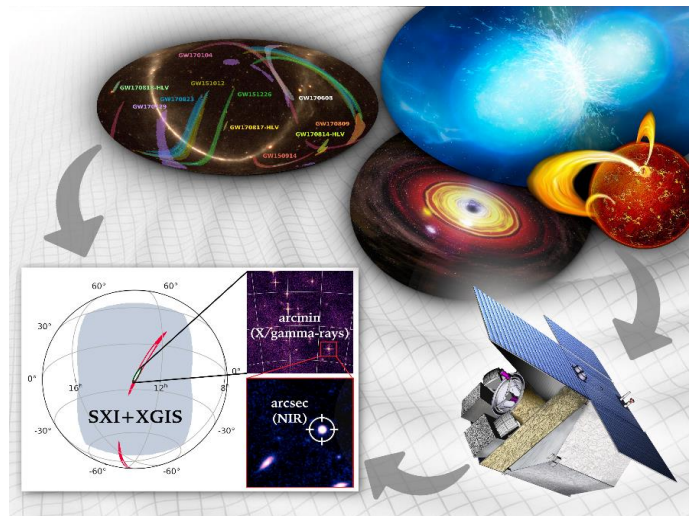

Figure 2 Examples of THESEUS capabilities for multi-messenger and time-domain astrophysics.

In this article we summarize the updated scientific requirements, mission concept and main expected performances, as resulting from the almost completed Phase A Study by ESA. For more details on the scientific instruments design and expected performances we refer to the accompanying papers included in this proceeding volume. As well, insights on the science case, scientific requirements and epected performances will be provided in a set of dedicated "white papers".

## 2. SCIENTIFIC REQUIREMENTS FOR THESEUS

The scientific goals coming from a full exploration of the early Universe require the detection, identification and characterization of several tens of long GRBs in the first billion years of the Universe (z > 6), in the 4 years of nominal mission lifetime of THESEUS (Figure 1). This performance will be almost 2 orders of magnitude better with respect to what has been achieved (7 GRBs at z>6) in the last 20 years with past and current GRB detectors (e.g., Swift/BAT, Fermi/GBM, Konus-WIND) combined with intensive follow-up programs with ground small (robotic) and large (e.g., VLT) telescopes. As supported by intensive simulations performed by us and other works in the literature, this breakthrough performance can be achieved by overcoming the current limitations through an extension of the GRB monitoring passband to the soft X-rays with an increase of at least 1 order of magnitude in sensitivity with respect to previously flown wide-field X-ray monitors, as well as a substantial improvement of the efficiency of counterpart detection, spectroscopy and redshift measurement through on-board prompt NIR follow-up observations. At the same time, the objectives on multi-messenger astrophysics and, more generally, time domain astronomy (Figure 2), require: a) a substantial advancement in the detection and localization, over a large (>2 sr) Field-of-View (FoV) of short GRBs as electromagnetic counterparts of GW signals coming from Neutron Stars (NS), and possibly NS-Black Hole (BH), mergers, as demonstrated in the case of GW170814; b) the capability of monitoring the high-energy sky with an unprecedented combination of sensitivity, location accuracy and FoV in the soft X-rays, as well as extending imaging capabilities up to the hard X-ray band and spectroscopic / timing up to the soft gamma-rays.

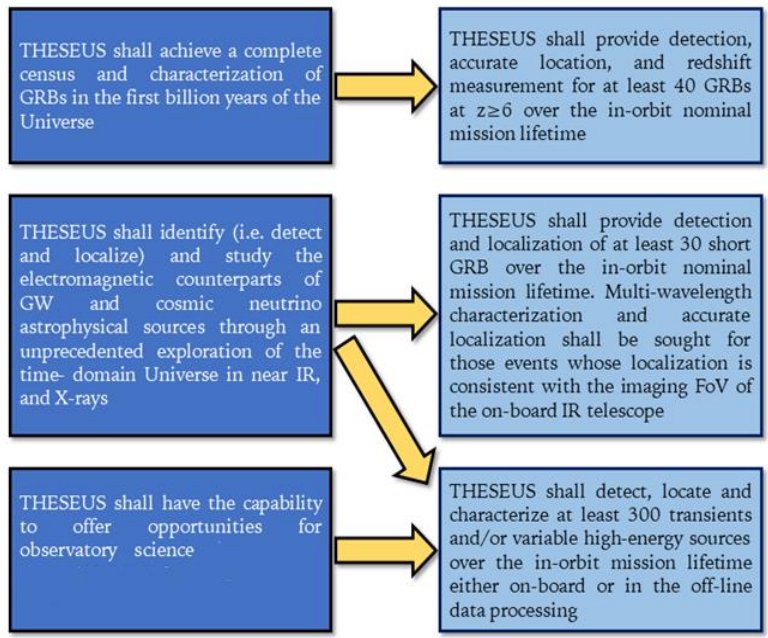

Figure 3 THESEUS main scientific requirements: flow-down from top level (left) to second level (right).

A summary of the main scientific requirements and their flow-down, as assessed by the THESEUS Science Study Team (TSST), is shown in Figure 3.

## 3. SCIENTIFIC PAYLOAD AND SPACECRAFT DESIGN

The performance can be best obtained by flying within a single mission a monitor based on the "lobster-eye" telescope technology, capable of focusing soft X-rays in the 0.3 – 5 keV energy band over a large FoV, with an innovative broad field of view hard X-ray / soft gamma-rays detection system, based on the joint use of Silicon Drift Detectors (SDD) and

CsI crystal scintillator bars, covering at least the same monitoring FoV as the lobster-eye telescopes and extending the energy band from a few keV up to several MeV. An on-board near-infrared telescope of the 0.5-1 m class is also needed, together with spacecraft fast slewing capability (5-10°/min), in order to provide prompt identification of the GRBs and other transients optical/IR counterparts, refinement of the positions down to ~arcsec precision, on-board redshift determination (thus enabling efficient follow-up with the largest ground and space observatories), and spectroscopy of the counterpart and of the host galaxy. The joint data from the whole scientific payloadwill characterize the luminosity, spectra and timing properties of transients over a broad energy band, enabling fundamental insights into their physics.

Based on the mission scientific requirements, and exploiting the unique heritage and worldwide leadership in the enabling technologies of the Consortium and ESA, the THESEUS payload will include the following scientific instruments:

- **Soft X-ray Imager** (SXI, 0.3 – 5 keV), a set of two X-ray telescopes based on MicroPore Optics (MPO) in "Lobser-eye" configuration and using large X-ray CMOS as detectors, covering 0.5 sr FOV with source location accuracy from 2 to 0.5 arcmin[4];

- **X/Gamma-rays Imaging Spectrometer** (XGIS, 2 keV – 10 MeV), a set of two X/gamma-rays coded mask telescopes using Silicon Drift detectors (SDD) coupled to CsI crystal scintillator detectors, with imaging and localization accuracy better than 15 arcmin in the 1 – 150 keV energy band and over a FOV of 2 sr, a few µs time resolution[5];

- **InfraRed Telescope** (IRT, 0.7 – 1.8 µ), an off-axis Korsch telescope with 0.7 m primary mirror granting a FOV of at least 15x15 arcmin and 1 arc sec source location accuracy, providing imaging capabilities in five NIR filters (I, Z, Y, J, H) and moderate (R~400 at 1.1µ) slit-less spectroscopy in 0.8 – 1.6µ over a 2x2 arc min FOV[6].

The main performance requirements for the THESEUS scientific instruments based on the flow-down form the higher level scientific requirements are reported in Table 1.

| Requirement | Value |
|---|---|
| SXI sensitivity (3σ) | $1.8 \times 10^{-11}$ erg cm$^{-2}$ s$^{-1}$ (0.3-5 keV, 1500 s) |
| | $10^{-10}$ erg cm$^{-2}$ s$^{-1}$ (0.3-5 keV, 100 s) |
| XGIS sensitivity (1s, 3σ) | $10^{-8}$ erg cm$^{-2}$ s$^{-1}$ (2-30 keV) |
| | $3 \times 10^{-8}$ erg cm$^{-2}$ s$^{-1}$ (30-150 keV) |
| | $2.7 \times 10^{-7}$ erg cm$^{-2}$ s$^{-1}$ (150 keV-1 MeV) |
| IRT sensitivity (imaging, End-of-Life, SNR=5, 150 seconds) | 20.9 (I), 20.7 (Z), 20.4 (Y), 20.7 (J), 20.8 (H) |
| SXI field-of-view (≤10% vignetting area) | 0.5 sr |
| XGIS field-of-view (>20% efficiency) | 2 sr (2-150 keV) |
| | 4 sr (≥150 keV) |
| IRT field-of-view | 15'x15' |
| Redshift accuracy (6≤z≤10) | ≤10% |
| IRT resolving power | ≥400 |
| XGIS background stability | ≤10% (over 10 minutes) |
| Field-of-Regard | ≥50% |
| Trigger broadcasting delay to ground-based networks | ≤30 seconds (65% of alerts) |
| | ≤20 minutes (95% of alerts) |
| SXI positional accuracy (0.3-5 keV, 99% c.l.) | ≤2 arcminutes |
| XGIS positional accuracy (2-150 keV, 90% c.l.) | ≤7 arcminutes (50% of the triggered sGRB) |
| | ≤15 arcminutes (90% of the triggered sGRB) |
| IRT positional accuracy (5σ detections) | ≤5 arcsecond (real-time) |
| | ≤1 arcsecond (post-processing) |

Table 1 Main requirements for the THESEUS intruments in-flight performances.

The key aspects of the design of the THESEUS spacecraft (SC) are driven by the VEGA-C vehicle capabilities, the requirements of the three different instruments and the on-board autonomous (i.e. without ground intervention) capabilities demanded by the concept of operations. In Figure 4 we show drawings of the possible THESEUS spacecraft design and scientific payload accommodation coming for the Phase A industrial studies. Both industrial teams developed a bus design based on 3-axis stabilised spacecraft with box shaped structure and divided into a Service Module and a Payload Module, with a central IRT telescope surrounded by the SXI and XGIS cameras. The AOCS is based on 4 star trackers optical heads configuration with 2 always in tracking, coarse rate sensor and sun sensors. The requird spacecraft agility is obtained through (3N+1R) Reaction Wheels, with magnetic torquers for quasi-continuous desaturation. The field of regard corresponds to 60% accessibility of the sky at any time. The spacecraft dry mass at launch is 1900 kg, including contingency (20% system margin).

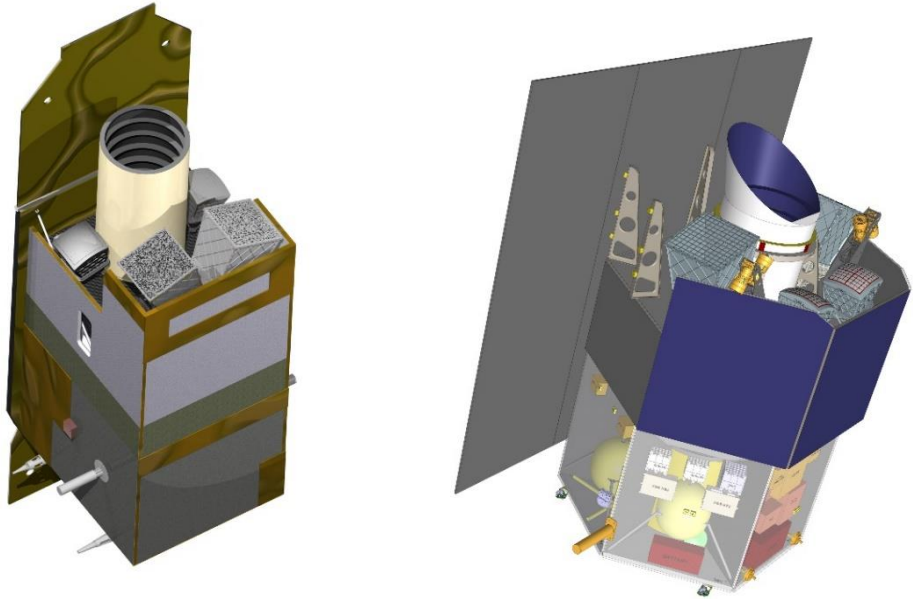

Figure 4 Schematic view of the spacecraft design from the industrial Phase A study by *Airbus Defence and Space (ADS*, left) and *Thales Alenia Space (TAS*, right).

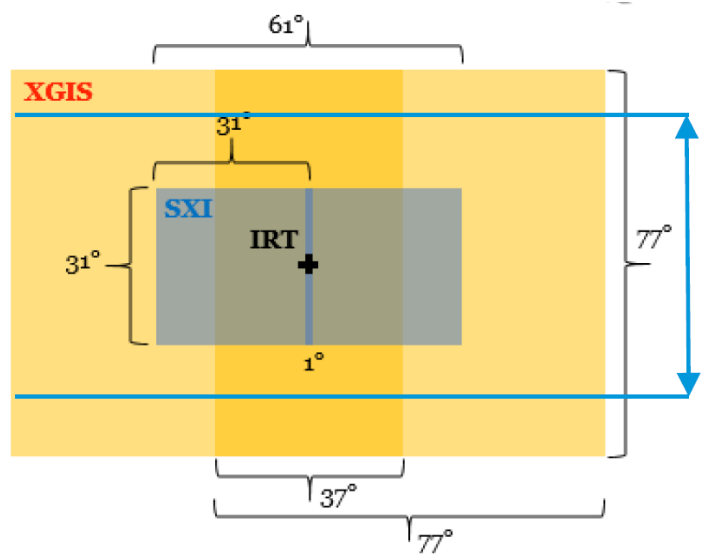

Figure 5 Relative alignment and FOVs of THESEUS instruments. (*ESA*)

From the programmatic point of view, the SXI is led by UK (with contributions by Belgium, Spain, Czech Republic. Ireland, The Netherlands and ESA), the XGIS is led by Italy (with contributions by Spain, Denmark and Poland) and the IRT is led by France (with contributions by Switzerland and ESA). The on-board Data Handling Units (DHU) system, capable of detecting, identifying and localizing likely transients in the SXI, XGIS and IRT FOVs, is responsibility of Germany, with contributions by Denmark and Poland.

## 4. MISSION PROFILE

Key features of the THESEUS mission profile will include the capability of promptly (within a few tens of seconds at most) transmitting to ground the trigger time and positions of GRBs (and other transients of interest) through the *Trigger Broadcasting Unit (TBU)* VHF transmitter, provided by Italy, and the *THESEUS Burst Alert Ground Segment (TBAGS)* network of VHF antennae, provided by France, as well as a spacecraft autonomous slewing capability >7°/min. The baseline launcher / orbit configuration is a launch with Vega-C to a low inclination (5.4°) Low Earth Orbit (LEO, 550-640 km altitude), which has the unique advantages of granting a low and stable background level in the high-energy instruments, allowing the exploitation of the Earth's magnetic field for spacecraft fast slewing and facilitating the prompt transmission of transient triggers and positions to the ground. The main ground station will be in Malindi (Kenya), provided by Italy. The mission nominal duration will be 4 years (corresponding to about three and half years of scientific operations), even though no technological issues preventing an extension by at least 2 more years have been identified.

The baseline *mission operation concept* (Figure 6) includes a survey mode, during which the monitors are chasing GRBs and other transients of interest, and, following a GRB (or transient of interest) trigger validated by the DHU system, burst acquisition and slew, follow-up and characterization modes. The *pointing strategy* during the survey mode will be such to maximize the combined efficiency of the sky monitoring by SXI and XGIS and that of the follow-up with the IRT. The technical feasibility of such pointing strategies has been demonstrated through a *Mission Observation Simulator*, which also demonstrated the overall compliance of the mission profile and instruments performance with the scientific requirements. Small deviations (of the order of a few degrees) from the survey mode pointing strategy will be possible so to point the IRT on sources of interest pre-selected through a *Guest Observer (GO) programme*. Once the main goals of the mission are achieved, accepted GO observations will also include pointings to sources or sky regions requiring much larger deviations from the nominal pointing strategy. Scientific modes also include external trigger (or TOO) mode, in which the IRT and monitors will be pointed to the direction of a GRB, transient or, e.g., to the error region of a GW or neutrino signal, provided by an external facility.

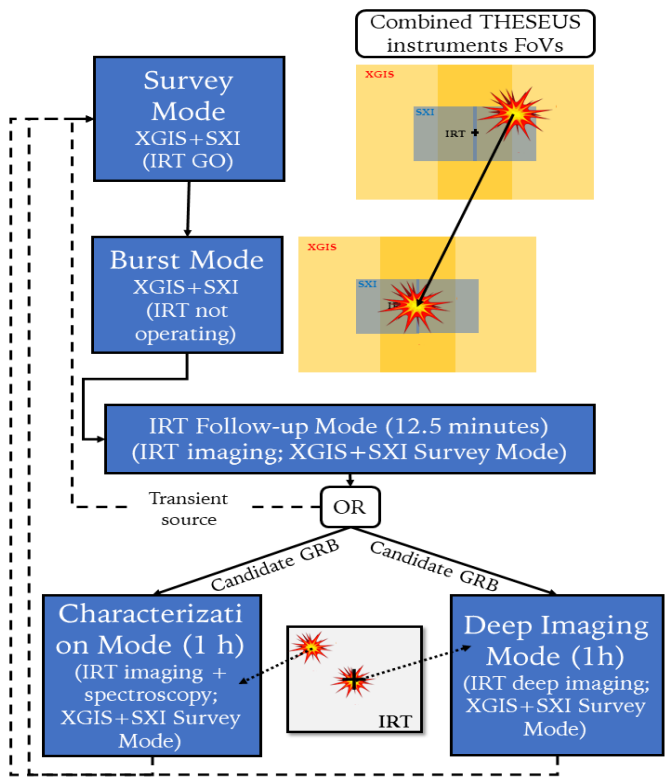

Figure 6 Scheme of the THESEUS scientific observation modes.

The Consortium Science Data Center (SDC) will be under the responsibility of Switzerland, with contributions from the other Consortium members. An overview of the THESEUS ground segment and organizational data flow is shown in Figure 7.

The **data policy** for THESEUS will maximize the participation of the international community and minimize cost and complexity of the management by the scientific ground segment. As in the case of the very successful Swift mission, the data will be reserved to the Instrument Teams only during the launch and Early Orbit Phase; data rights will be extended to scientists of the whole consortium during the Performance Verification Phase; and after that, data will be public as soon as they are processed. ESA and the Consortium will do their best efforts for supporting the scientific community in the full exploitation of THESEUS data.

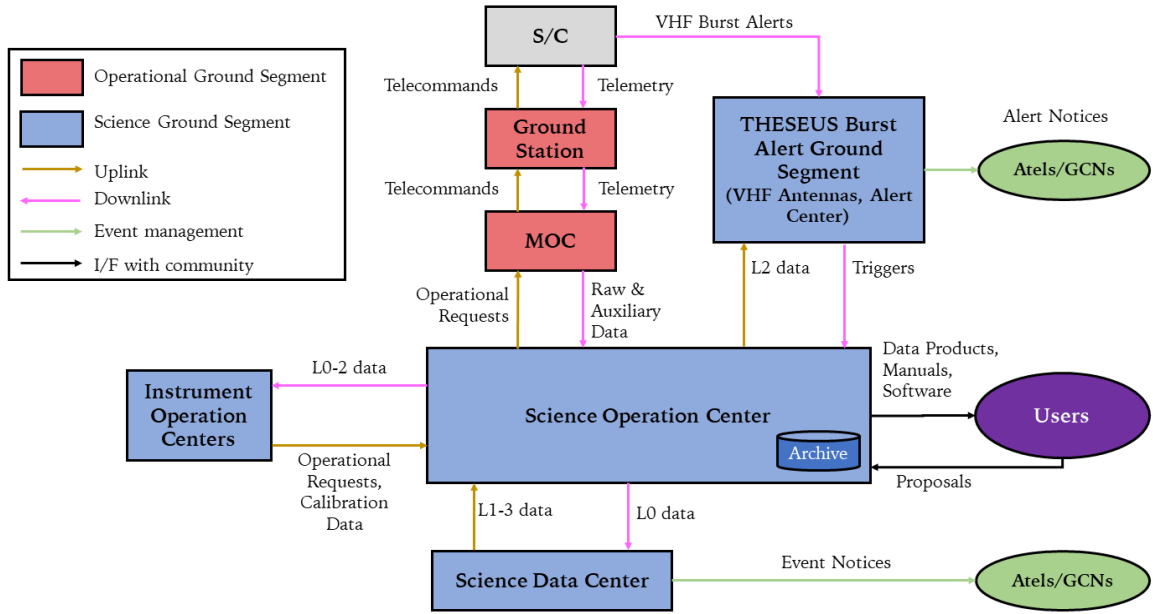

Figure 7: Overview of the THESEUS ground segment organization and data flow.

## 5. EXPECTED PERFORMANCES

The top-level THESEUS scientific requirements described in Section 2 and Figure 3 have been verified through a Mission Observation Simulator (MOS) developed by ESA with contributions by the Consortium. A realistic observational sequence of the THESEUS spacecraft according to the operational modes described in Section 4 and Figure 6 was simulated considering all observational constraints (Earth occultations and eclipses, South Atlantic Anomaly passages) in response to a random set of short and long GRB triggers as per a GRB population model developed by the Consortium[7]. External triggers (three per month as per science requirement) as well as estimated false alert rates (three and one per week for the SXI and XGIS, respectively, as per science requirements augmented in the case of the SXI) were injected randomly in the simulations to estimate the associated inefficiencies. In order to achieve sufficient statistics, the results of 40 simulations of 4-years THESEUS nominal operations (corresponding to 3.45 years of science observations) were merged together.

Several possible pointing strategies have been evaluated, differing by the distribution of pointing directions during the Survey Mode. All pointing strategies must be compliant by design to a Solar Aspect Angle larger than 60 degrees. A trade-off was carried out during Phase A between a fixed Survey Mode pointing to the Ecliptic Poles (switching between the North and the South Pole every six months; "ECP" hereafter), and pointing strategies where two tilts per orbit by 30 or 60 degrees were allowed (the latter one will be identified as "DYN" hereafter). The former strategy maximizes the absolute rate of GRB triggers detected by the high-energy instruments; the latter strategies maximize the rate of THESEUS-detected triggers that can be follow-up by ground-based telescopes (typically located within ±30 degrees latitude). The SXI exposure map and normalized distributions of triggers' latitudes corresponding to the DYN pointing strategy are shown in Figure 8.

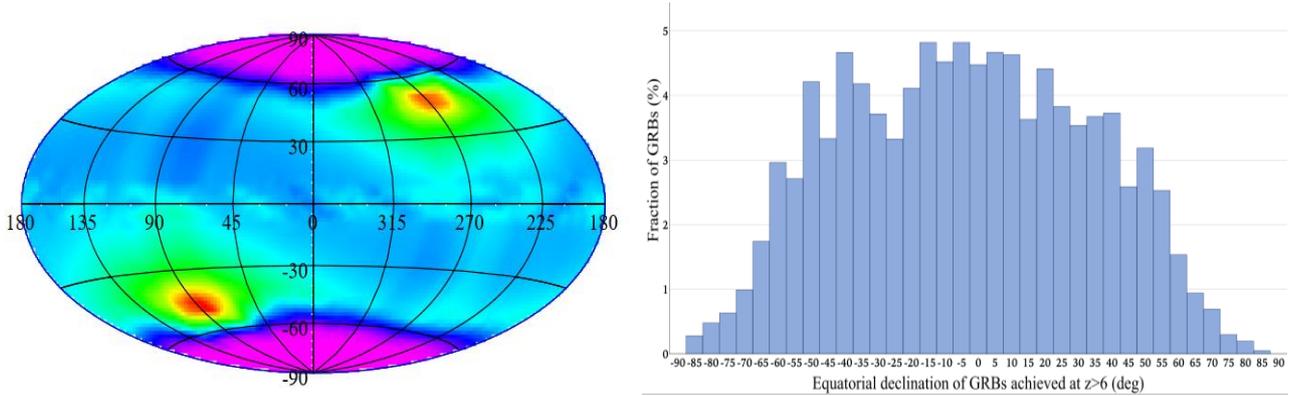

Figure 8 –SXI exposure maps in equatorial coordinates (left) and distribution in declination of z>6 GRBs for the DYN pointing strategy during Survey Mode. (*S. Mereghetti and THESEUS Consortium*)

About 90% (50%) of the GRB triggers detected with the DYN strategy have a latitude lower than 55 (30) degrees, favourable to ground-based follow-up observations, against 30% (4%) for the EP strategy. The sky coverage of the survey with a DYN strategy is correspondingly more homogeneous than with the ECP strategy. The mission analysis confirmed that the basic science objectives of the mission can be achieved with both pointing strategies. The ECP yields more than 45 long GRBs at z>6 and at least 40 short GRBs during the 4- year nominal operations. The baseline DYN strategy yields about 15% less short GRBs, and about 5% more long GRBs. The combination of more homogeneous survey sky coverage, higher probability of ground-based follow-up observations of THESEUS-detected triggers, and comparable trigger detection efficiencies leads to the DYN pointing scenario to be chosen as the baseline during Phase A.

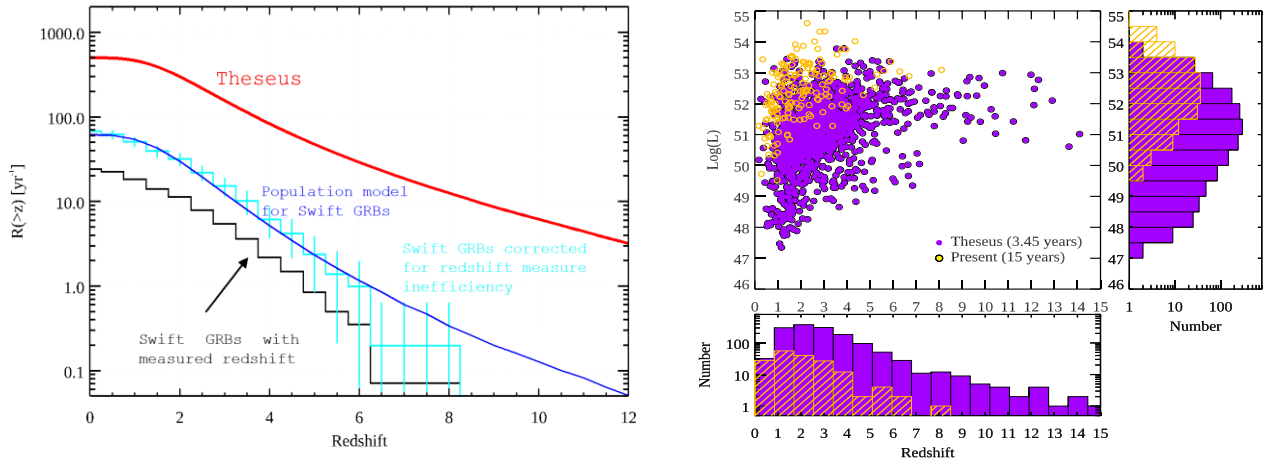

Figure 9 – Left Panel: expected detection rate of long GRBs by THESEUS (red curve) compared with the detected rate observed by Swift (grey curve). The cyan curve represent the observed Swift distribution corrected for redshift inefficiencies. The blue curve represents the model used to simulate the expected THESEUS distribution that fits well the Swift observational data. Right Panel: Distribution of long GRBs in the luminosity versus redshift plane now (yellow points and hatched histogram) and after the nominal operation life of THESEUS (purple points and hatched histogram). (*G. Ghirlanda and THESEUS Consortium*)

The outstanding performances of THESEUS in terms of detection, accurate location and identification of high-z GRBs compared to what has been obtained with past and current space facilities and ground observatories, already anticipated and put in the in the cosmological context in Figure 1, are shown in more details in Figure 9, where the DYN pointing strategy has been assumed. With an estimated rate of 12-14 GRBs at z>6 identified per year, THESEUS will perform about

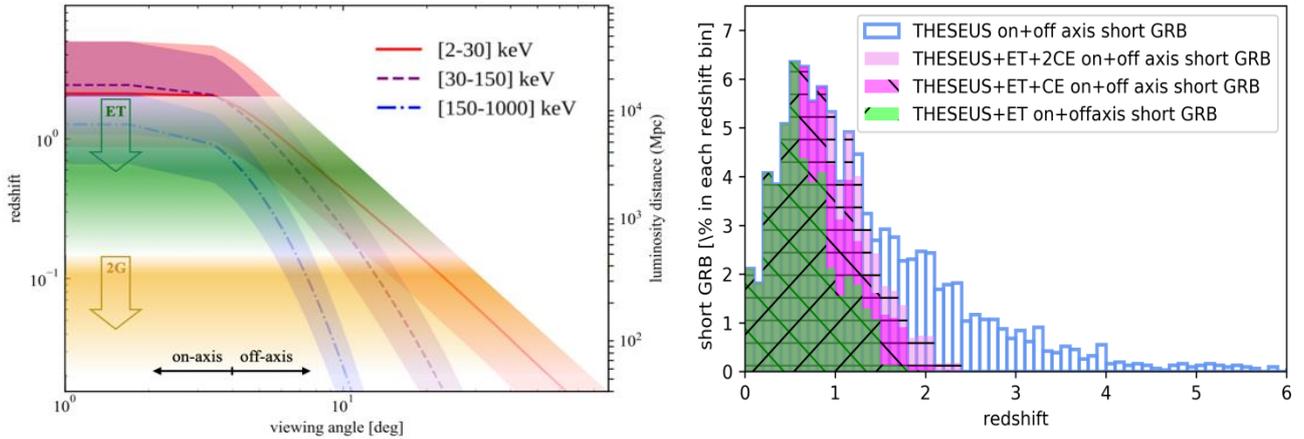

Figure 10 Left: The pink, violet and blue stripes indicate the maximum distance/redshift for detecting with THESEUS/XGIS the prompt emission of a 170817A-like short GRB versus the viewing angle. Green and orange shades indicate the ET and 2G distance reach for NS-NS mergers, respectively. At 330 Mpc (typical 2G distance reach for random inclinations), for instance, THESEUS can detect up to a viewing angle of ~20°-40° (2-30 keV band), providing >25-100 times more detections with respect to a viewing angle within the jet core (<4°). (*O. Salafia and THESEUS Consortium*). Right: The redshift distribution of well-localized on-axis and off-axis short GRBs (blue) from XGIS and SXI and those detected also with IRT (25%, indigo) Joint short GRB+GW detections are obtained by considering, at each redshift, the GW detection efficiency for on-axis NS-NS mergers with ET. Including off-axis events not only increases the total number of THESEUS short GRBs, but also boosts the fraction of events with a joint EM+GW detection, leading to ~47% for ET, ~64% for ET+CE and ~74% for ET+2CE. *(S. Grimm, M. Branchesi, J. Harms and THESEUS Consortium).*

40 times better than past and current worldwide facilities all together, which provided about 7 of these events in more than 20 years (corresponding to about 0.35 events per year).

As outlined in Section 1, among the outstanding capabilities of THESEUS for multi-messenger astrophysics, the most promising is the detection and accurate location of short GRBs, and possibly of their associated KN emission, produced by NS-NS or NS-BH mergers, being up to now the only e.m. counterparts to a GW signal ever detected (GW170817). Under this respect, as shown in Figure 10, the capability by THESEUS monitors of extending the GRB prompt emission detection and characterization down to soft X-rays (whereas current detectors like Swift/BAT, Fermi/GBM or Konus-WIND are limited to >10 keV) will allow to improve substantially the number of coincident detections with GW detectors expected to operate in the '30s with respect to the current scenario.

Finally, Figure 11 gives a flavour of the more general tremendous capabilities of THESEUS for time-domain astronomy. The massive grasp high sensitivity and wide spectral coverage of the THESEUS high-energy instruments offer a great opportunity to study the variability properties of several classes of sources on timescales from seconds to years and to promptly trigger observations with other facilities when new sources or interesting phenomena are discovered. Currently, no X-ray all-sky monitors are planned for the 2030 timeframe; there will be no other instruments that will have the sky coverage to detect outbursts and trigger observations with other ground and space facilities. THESEUS will provide triggers from the discovery of new sources, and also from the detection of peculiar states in known sources that will be regularly monitored. Additionally, the use of the IRT for selected sources, through the implementation of a Guest Observer Program (see section 2.6), further broadens the scientific scope by allowing for unique multi-wavelength monitoring.

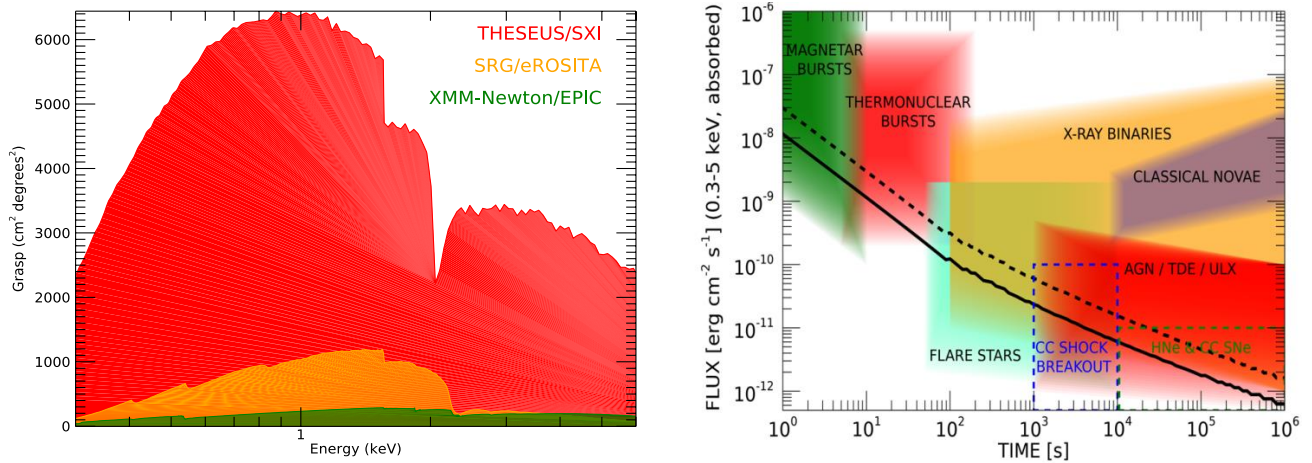

Figure 11 – Left panel: SXI grasp (red area) as a function of energy compared to that of the X-ray survey mission SRG/eROSITA (orange), and XMM-Newton/EPIC (green). Credit for the XMM-Newton and eROSITA data: A. Merloni (MPE). Right Panel: Typical variability time scales and soft X-ray fluxes of different classes of transient / variable sources compared to the SXI sensitivity for a power law spectrum with photon index 2 and neutral Hydrogen equivalent column density NH $=5\times10^{20}$ cm$^{-2}$ (solid) and $10^{22}$ cm$^{-2}$ (dashed). *(M. Guainazzi, S. Mereghetti and THESEUS Consortium)*

## 6. GUEST OBSERVATORY SCIENCE AND SYNERGIES

A Guest Observer programme, including target of opportunity (TOO) observations, will further expand the scientific return of the mission. Indeed, in addition to the main core goals and expected performances, summarized in previous sections, THESEUS is going to provide a very special opportunity for agile IR and X-ray observations of a wide range of targets, from asteroids, to exoplanets, to AGNs. Indeed, space-borne, sensitive NIR spectroscopy is an extremely useful capability, and wide-field sensitive X-ray monitoring can identify changes and priorities for follow-up studies.

While in Survey Mode, the IRT, SXI and XGIS will be continuously gathering data, with IRT pointed at a specific target within a few degrees from the baseline optimal pointing strategy. Hundreds of thousands of suitable targets are already known, and eROSITA, Euclid, the Vera Rubin Observatory (VRO, also known as LSST) and SKA are deepening and extending the range of relevant catalogues. Many targets for THESEUS as an observatory have a time-domain aspect. A space-based infrared, and X-ray spectroscopic facility will be attractive to a wide range of investigators, and address important questions in a plethora of scientific areas. While less powerful than JWST and ATHENA, the chance to use THESEUS to observe substantial samples of interesting sources, both known and newly-discovered, to appropriate depths and cadences, while the mission is searching for GRBs, provides opportunities for additional science, for a user community interested in scales all the way from the Solar System to distant AGN can provide abundant desired targets for THESEUS as an observatory.

The co-alignment of the IRT with the part of the SXI FoV where the two units overlap ensures that the best X-ray spectra/limits will be obtained alongside every IR imaging/spectral target. For instance, an IR spectrograph in space is able to investigate a range of cometary emission and absorption features, without being restricted to specific atmospheric bands, and with full access to all water and ice features, impossible from the ground. Several tens of comets per year are likely to be observable as they pass through the inner solar system, evolving through their approach to and recession from perihelion. IR spectra of large samples of stars with transiting planets can be obtained by THESEUS. By 2030, tens of thousands of transiting planets will be known, spread widely over the sky, and with well-determined transit times, which can be scheduled well in advance to search for potential atmospheric signatures in IR absorption spectroscopy. IRT is more sensitive than ARIEL, and so carefully chosen extended planetary transit observations can be made for known targets, and for a substantial number of transiting planetary targets can be included in the observatory science target catalogue. X-ray binaries and flaring stars can be discovered as bright X-ray and IR spectral targets by THESEUS,

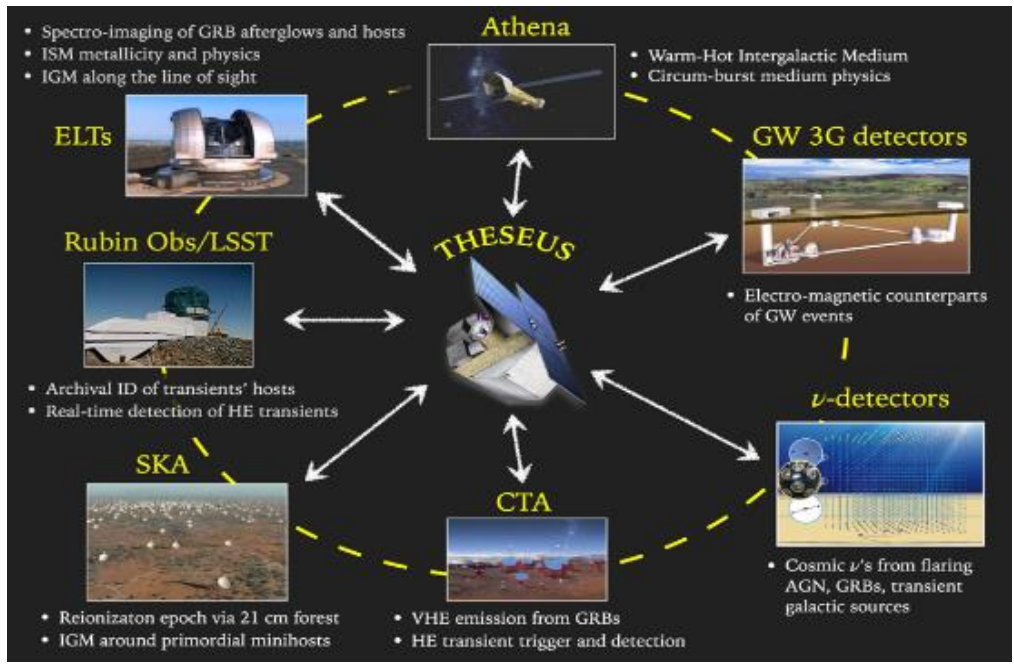

Figure 12 THESEUS will work in synergy on a number of themes (bullets) with major multi-messenger facilities in the 30ties and will provide targets and triggers for follow-up observations with several of these facilities. (*P. Rosati and THESEUS Consortium*)

or highlighted by other observatories, and then confirmed and studied using THESEUS's spectroscopic capabilities. Found predominantly in the Galactic Plane, many hundreds of bright events will occur during the nominal mission.

A prompt spectroscopic IR survey for supernovae that are found taking place out to several 10s of Mpc, unencumbered by atmospheric effects, is likely to remain attractive beyond 2030, and will help to resolve remaining questions about the impact of environment and metallicity on the nature of supernovae and their reliability as standard candles. Without sensitive IR spectroscopy, these questions might not be resolved. The availability of the full spectral window is particularly helpful for observations of emission-line galaxies and AGN, for which key diagnostic lines are redshifted out of the optical band from the ground at redshifts z~0.7. Even ELTs cannot beat the atmosphere, and huge candidate samples will be catalogued over large fractions of the sky, colour-selected from VRO surveys, in concert with the coverage of eROSITA, SKA and WISE. IRT will enable H☐ spectral surveys of interesting classes of the most luminous galaxies and AGN all the way to z~2-3. Furthermore, the THESEUS mission will provide a useful time baseline out to several years, to see potential changes in the appearance of AGN spectra, and to confirm any changes by revisiting selected examples. There are samples of many thousands of AGN, selected in a variety of ways from Rubin, WISE and SKA, that would be interesting to investigate and possibly monitor with THESEUS. While it will be impossible to include more than a few thousand galaxies and AGN in a spectral monitoring programme, the results of combining the wide-area data from VRO and WISE in the optical and IR, and with eROSITA in the X-ray, with the serendipitous wide-area coverage of SXI and XGIS, will allow new insight into the X-ray variability of large samples of AGN.

Finally, as already detailed in Section 1 and is illustrated in Figure 12, for all aspects of its core and observatory science THESEUS will show a great synergy with future large observing facilities in both the electromagnetic (like, e.g., LSST, ELT, TMT in the optical, SKA in the radio, CTA in VHE and ATHENA in X-rays) and multi-messenger (e.g., aLIGO-plus, aVirgo-plus, KAGRA+; Einstein Telescope, Cosmic Explorer, for GW, and Km3NET and IceCube-Gen2 for neutrinos). The combination and coordination of THESEUS with these multi-wavelength, multi-messenger facilities expected to be operating in the thirties will open new avenues of exploration in many areas of astrophysics, cosmology and fundamental physics, thus adding considerable strength to the overall scientific impact of THESEUS and allowing them to fully exploit their scientific capabilities.

## ACKNOWLEDGEMENTS

We acknowledge relevant contributions to the text and figures reported in this article by the THESEUS ESA Study Team (TEST) and Science Study Team (TSST), as well as from the THESEUS Consortium Editorial Board (TEB) and scientific Working Groups. See http://www.isdc.unige.ch/theseus for complete members lists of these teams. Special thanks for their direct or indirect contributions to the content of this article go to S. Basa, A. Blain, M. Branchesi, C. Labanti, F. Frontera, G. Ghirlanda, S. Grimm, M. Guainazzi, I. Hutchinson, H. Lerman, S. Mereghetti, J. Osborne, F. Pinsard, P. Rosati, G. Stratta, N. Tanvir, C. Tenzer, R. Campana, R. Casesa, V. Da Ronco, Y. Evangelista, P. Hedderman, G. Parissenti, A. Rocchi, O. Salafia. Thales Alenia Spazio and Airbus Defence and Space. In addition to ESA, the THESEUS Phase A Study in the framework of the ESA Cosmic Vision M5 programme is supported by the National Agencies and Space Organizations of the Consortium member states (Italy, UK, France, Germany, Switzerland, Spain, Denmark, Poland, Belgium, Czech Republic, The Netherlands, Ireland and Slovenia), either directly or through the PRODEX programme. Further support by ESA to technological R&D activities of the Consortium is provided through the M5/NPMC programme. LA acknowledges support by ASI-INAF Agreement n. 2018-29-HH.0. D.G. acknowledges financial support by LabEx UnivEarthS (ANR-10-LABX-0023 and ANR-18-IDEX-0001). The Phase A study of THESEUS is also supported by the AHEAD2020 project funded by UE through H2020-INFRAIA-2018-2020.